\documentclass[aps,preprint,nofootinbib,superscriptaddress]{revtex4}

\usepackage{amssymb}

\usepackage{epsfig}
\usepackage[bookmarksnumbered,bookmarksopen,colorlinks,citecolor=blue,linkcolor=blue]{hyperref}

\begin{document}

\title{ Microscopic dynamics simulations of multi-nucleon transfer in $^{86}$Kr+$^{64}$Ni at 25 MeV/nucleon }

\author{Hong Yao}
\affiliation{ Department of Physics,
Guangxi Normal University, Guilin 541004, People's Republic of
China }

\author{Ning Wang}
\email{wangning@gxnu.edu.cn}\affiliation{ Department of Physics,
Guangxi Normal University, Guilin 541004, People's Republic of
China }

\begin{abstract}

Multi-nucleon transfer in $^{86}$Kr+$^{64}$Ni at an incident energy of 25 MeV/nucleon is for the first time investigated with a microscopic dynamics model: improved quantum molecular dynamics (ImQMD) model. The measured isotope distributions are reasonably well reproduced by using the ImQMD model together with a statistical code (GEMINI) for describing the secondary decay of fragments. The reaction mechanism is explored with the microscopic dynamics simulations from central to peripheral collisions. In central collisions there exists a strong competition among fusion, deep-inelastic scattering and multi-fragmentation at such an incident energy. In semi-peripheral collisions, binary scattering together with nucleon transfer is dominant, and the probability of elastic+inelastic scattering events  increases rapidly with impact parameter in peripheral collisions and approaches to one when $b>14$ fm. The mass-TKE distribution in central collisions due to the competition is quite different from those in peripheral collisions and the distribution of total kinetic energy loss (TKEL) for binary events with nucleon transfer is much more broader than those without transfer.

\end{abstract}
\maketitle

\begin{center}
\textbf{I. INTRODUCTION}
\end{center}

Structure of neutron-rich nuclei, such as their masses and neutron-skin thicknesses, is of great importance for investigating nuclear symmetry energy \cite{Li14,Stein,Khan12,Tsang09,Cent09,Wang13}, nucleus-nucleus collisions \cite{OSm,Timm,Zhang10,Setf06}, and r-process in astrophysics \cite{Temis15,Mum16,LZ12}. Unfortunately, the number of observed extremely neutron-rich nuclides is very limited and about 4000 masses of neutron-rich nuclides in nuclear landscape are still unmeasured, due to that heavy-ion fusion reactions \cite{OPb2,Hof00,Ogan15,Sob,Gup05,Wang09,Wangbin} with stable beams as the traditional approach to synthesize new heavy nuclei cannot produce these extremely neutron-rich nuclei. It is therefore necessary to search for alternative way to extend the nuclear landscape. The synthesis of these neutron-rich nuclides through multi-fragmentation, deep inelastic scattering and quasi-fission are of exceptional importance to advance our understanding of nuclear structure at the extreme isospin \cite{Shen87,Ober14,Zag11,Sou02,Sou03,Heinz14}. In addition to the formation of neutron-rich fragments from fission of actinides or projectile fragmentation, multi-nucleon transfer process is also helpful to produce neutron-rich heavy nuclei.

The production of neutron-rich nuclei in multi-nucleon transfer (MNT) between two heavy-ions has attracted a lot of attention in recent years. For example,
Watanabe et al. measured the absolute cross sections for neutron-rich isotopes around and beyond the neutron shell $N =126$ formed in MNT in the $^{136}$Xe+$^{198}$Pt system at $\sim$8 MeV/nucleon \cite{Wata}. Barrett et al. investigated experimentally the transfer reaction in the $^{136}$Xe+$^{208}$Pb system at about 5.5 MeV/nucleon \cite{Barr}, and Li et al. analyzed this reaction with the improved quantum molecular dynamics (ImQMD) model \cite{Li16}. Wang and Guo investigated MNT process in $^{154}$Sm+$^{160}$Gd at $\sim$5.6 MeV/nucleon with two different microscopic dynamics approaches:  ImQMD model and time dependent Hartree-Fock (TDHF) theory \cite{SmGd}. No fusion was observed from both models for this reaction, whereas more than 40 extremely neutron-rich unmeasured nuclei with $58 \le Z \le 76$ are observed and the predicted production cross sections are at the order of $\mu$b to mb. Corradi et al. studied the MNT reactions in  $^{64}$Ni+ $^{238}$U at about 6.1 MeV/nucleon and confirmed that a clear experimental distinction can be made between the collisions in the grazing (quasielastic and deep-inelastic) regime and in a more complex one (quasifission) \cite{NiU99}, and this reaction was also described with TDHF very recently \cite{Sek16}. Experimentally, Szilner et al. investigated the MNT processes in $^{40}$Ca+$^{208}$Pb \cite{CaPb05} and Nishio et al. investigated the fusion probabilities in the reactions $^{40,48}$Ca+ $^{238}$U \cite{CaU12} at energies around the Coulomb barrier. The MNT process in collisions between two massive nuclei such as $^{238}$U+$^{238}$U at near-barrier energy was also extensively investigated both experimentally and theoretically \cite{UU09,UU13,UUTDHF,UUWang,UUZhao}, and the production of unknown neutron-rich isotope was predicted very recently with the ImQMD model \cite{Zhao16}.

In addition to the reactions at energies around $5 \sim 8$ MeV/nucleon, a large enhancement in the production of neutron-rich projectile residues was observed in the reaction $^{86}$Kr+$^{64}$Ni at 25 MeV/nucleon \cite{Sou02,Sou03}. At an incident energy of 25 MeV/nucleon which is much higher than the Coulomb barrier, the competition among fusion, quasi-elastic scattering, deep inelastic scattering and multi-fragmentation could be expected. It is therefore interesting to investigate the competition and the influence of nucleon transfer on kinetic energy distribution of fragments and reaction time.

To investigate the MNT processes theoretically, the semiclassical GRAZING model \cite{Win94,Love15} has been developed with great successes \cite{Corr09}. In addition, a dynamical model based on Langevin-type equations \cite{Zag08} and the dinuclear system (DNS) model \cite{Ada10,Ada10a,Wangnan} are also proposed to describe the nucleon transfer process. To understand the dynamical nucleon transfer process in fusion and deep inelastic scattering reactions more microscopically, some microscopical dynamics models, such as  TDHF \cite{Naka05,Maru06,Guo07,Sim12,Sim14} and the ImQMD model   \cite{ImQMD2002,ImQMD2004,ImQMD2010,ImQMD2014} have also been developed. In the ImQMD model, the standard Skyrme force with the omission of spin-orbit term is adopted for describing not only the bulk properties but also the surface properties of nuclei. Simultaneously, the Fermi constraint which was previously proposed by Papa et al. in the CoMD model \cite{constrain} and improved very recently in Refs. \cite{Wang15,Wang16} is used to maintain the fermionic feature of the nuclear system. The ImQMD model allows to investigate the fluctuations and fragment formation during a heavy-ion collision in a consistent $N$-body treatment, through event-by-event simulations. One of the aims in this work is to investigate the competition among fusion, quasi-elastic scattering, deep inelastic scattering, ternary breakup and multi-fragmentation in $^{86}$Kr+$^{64}$Ni at 25 MeV/nucleon with the microscopic dynamics model.

The structure of this paper is as follows: In sec. II, the framework of ImQMD and details in simulating $^{86}$Kr+$^{64}$Ni will be introduced. In sec. III, the calculated results about $^{86}$Kr+$^{64}$Ni at an incident energy of 25 MeV/nucleon will be presented and analyzed. Finally a summary will be given in Sec. IV.

\begin{center}
\textbf{ II. Theoretical Framework and Details in Simulations}\\
\end{center}

In ImQMD-v2.2 simulations \cite{Wang16,SmGd}, each
nucleon is represented by a coherent state of a Gaussian wave
packet
\begin{equation}
\phi _{i}(\mathbf{r})=\frac{1}{(2\pi \sigma _{r}^{2})^{3/4}}\exp \left [-\frac{(%
\mathbf{r}-\mathbf{r}_i)^{2}}{4\sigma _{r}^{2}} +\frac{i}{\hbar} \mathbf{r} \cdot \mathbf{p}_i \right ],
\end{equation}
where $\mathbf{r}_{i}$ and $\mathbf{p}_{i}$ are the centers of the $i$-th
wave packet in the coordinate and momentum space, respectively.
$\sigma _{r}$ represents the spatial spread of the wave packet.
The total $N$-body wave function is assumed to be the direct product
of these coherent states. The anti-symmetrization effects are additionally simulated by introducing the Fermi constraint. Through a Wigner transformation, the one-body  phase space distribution function and the density distribution function $\rho$ of a system
\begin{equation}
\rho(\mathbf{r})=\sum_i{\frac{1}{(2\pi \sigma_r^2)^{3/2}}\exp
\left [-\frac{(\mathbf{r}-\mathbf{r}_i)^2}{2\sigma_r^2} \right ]},
\end{equation}
are obtained. The propagation of nucleons is governed by the self-consistently generated mean-field,
\begin{equation}
\mathbf{\dot{r}}_i=\frac{\partial H}{\partial \mathbf{p}_i}, \; \;
\mathbf{\dot{p}}_i=-\frac{\partial H}{\partial \mathbf{r}_i},
\end{equation}
and the momentum re-distribution in the Fermi constraint. Euler algorithm is adopted to compute new positions and momenta at time $t+ \Delta t$. The time step in the ImQMD calculations is set as $\Delta t=1$ fm/c. The Hamiltonian $H$ consists of the kinetic energy and the effective interaction potential energy which is based on the Skyrme EDF by neglecting the spin-orbit term.

We simulate $^{86}$Kr+$^{64}$Ni at 25 MeV/nucleon from central collisions to very peripheral collisions (impact parameter $b=1\sim14$ fm with $\Delta b=1$ fm). For each impact parameter we create 100,000 events and the ImQMD simulations are performed till $t=1000$ fm/c. The production cross sections of isotopes are finally predicted by using the ImQMD model together with a statistical code (GEMINI \cite{Char88}) for describing the secondary decay of fragments.

\begin{center}
\textbf{III. Results and Analysis}
\end{center}

\begin{figure}
\includegraphics[angle=0,width=0.9 \textwidth]{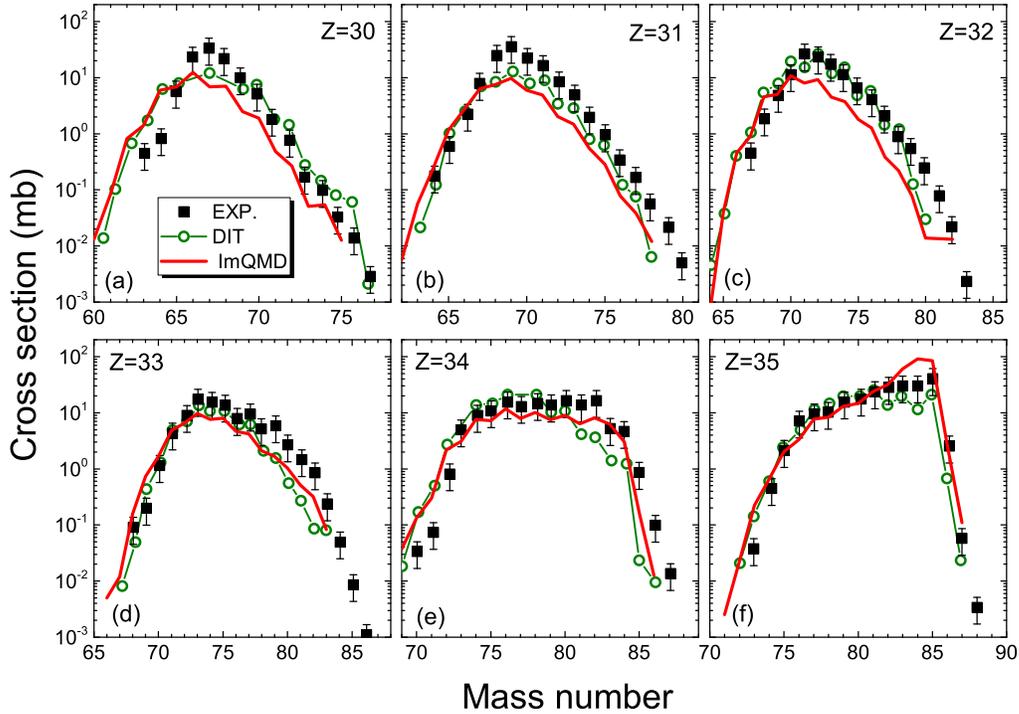}
\caption{(Color online) Isotopic distribution for elements $Z=30 \sim 35$ in $^{86}$Kr+$^{64}$Ni at 25 MeV/nucleon. The squares and the circles denote the experimental data and the predictions of DIT code taken from Ref.\cite{Sou02}, respectively.  The solid curves denote the results of ImQMD together with  GEMINI for describing the secondary decay of fragments.}
\end{figure}

We firstly check the ImQMD model for describing the mass distributions of elements $Z=30$ to $Z=35$ in the reaction of $^{86}$Kr+$^{64}$Ni at 25 MeV/nucleon.
Fig. 1 shows the comparison of the production cross section of nuclei with $Z=30 \sim 35$ in this reaction. The squares denote the experimental data taken from Ref.\cite{Sou02}. The circles (taken from \cite{Sou02}) denote the predictions of deep inelastic transfer (DIT) code of Tassan-Got and Stephan \cite{Tas91}. The solid curves denote the results of ImQMD calculations together with the code GEMINI for describing the secondary decay of fragments. It should be pointed out that the predicted fragment yield distributions from the ImQMD simulations are filtered with an angular range $10^\circ \sim 27^\circ$ in the laboratory frame considering the angular opening of MARS and an azimuthal angular factor of 10 introduced in the experiment as the corrections for the measured yields \cite{Sou03}. We find that the measured isotope distributions can be reasonably well reproduced by using the ImQMD+GEMINI calculations. The discrepancies from the data are within one order of magnitude in general, which is comparable with the predictions of DIT code. The version v2.2 of the ImQMD model used in the calculations is as the same as that described in Refs.\cite{SmGd,Wang16} and the parameter set IQ3a \cite{ImQMD2014} is adopted for the description of the mean field. In the description of the secondary decay with GEMINI, the excitation energy of an excited fragment is obtained as
the total energy of the fragment in the body frame with the corresponding ground-state binding energy being subtracted. The GEMINI parameters are chosen as following: The level density is taken as Grimes case B modified form ({\it aden\_type}=$-$23),
 all asymmetric divisions is considered in fission mode ({\it imf\_option}=2),
 the particle with $Z\le 5$ is treated in light particle evaporation ({\it Z\_imf\_min}=5).
 The masses of unmeasured nuclei are taken from the predictions of Weizs\"acker-Skyrme (WS4) mass model \cite{Wang14}. The other parameters are taken as the default value given by example in GEMINI document.

\begin{figure}
\includegraphics[angle=0,width=0.9 \textwidth]{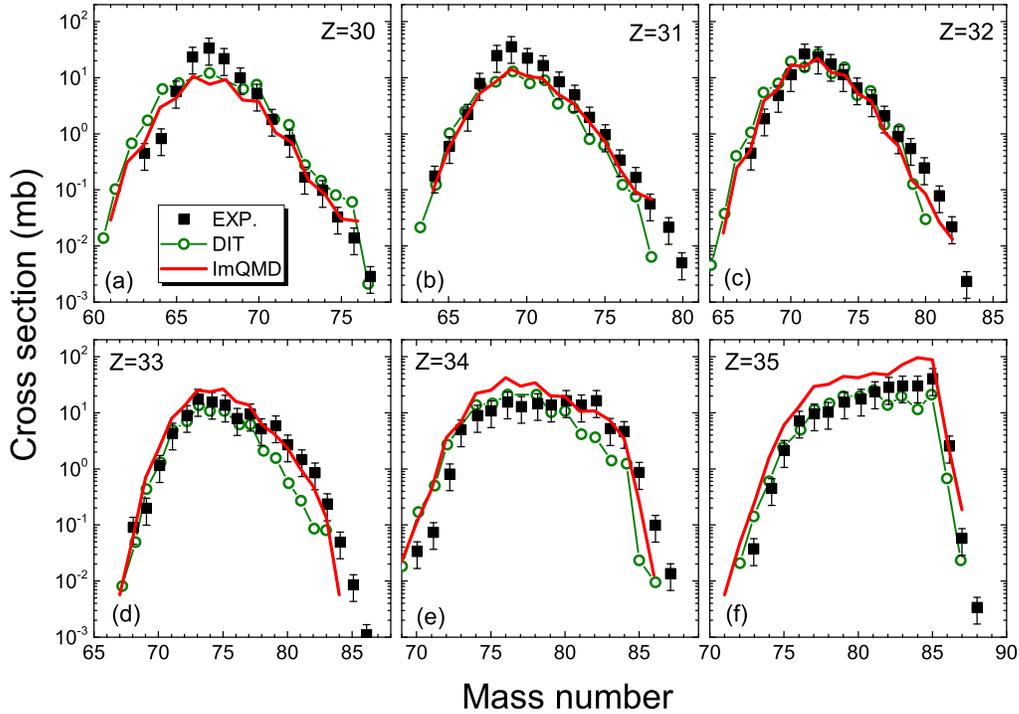}
\caption{(Color online) The same as Fig. 1, but filtered with an impact parameter range $8 <  b < 14$ in the ImQMD calculations. }
\end{figure}

It is expected that the quasi-elastic components of the multi-nucleon transfer process in $^{86}$Kr+$^{64}$Ni system at 25 MeV/nucleon mainly appears in the peripheral collisions. Therefore, in the DIT calculations the stochastic nucleon exchange was assumed for the orbital angular momentum range $\ell = 100 \sim 520$ and the events corresponding to trajectories in which the projectile-target overlap exceeded 3 fm were rejected \cite{Sou02}. In other words, the minimum impact parameter $b_{\rm min}$ is assumed and the DIT calculations are performed from semi-peripheral to very peripheral collisions. Here, the touching distance between two nuclei is estimated by
\begin{equation}
R_{12} = R_{ch}^{(1)}+R_{ch}^{(2)}+ \Delta R_{np}.
\end{equation}
Where, $R_{ch}$  denotes the charge radius of a nucleus, which can be well described by using the formula proposed by Wang and Li \cite{Radii}. Here $\Delta R_{np} \simeq 0.6 $ is taken to consider the neutron-skin and surface diffuseness effects between two neutron-rich nuclei. For $^{86}$Kr+$^{64}$Ni, the touching distance is $R_{12}\simeq 11$ fm. The impact parameters adopted in the DIT calculations are therefore about $b\approx R_{12}\pm 3 $ fm.

It is interesting to compare the model predictions from two different filters: the filter from angular range $10^\circ \sim 27^\circ$ and the one from impact parameter range $b_{\rm min} < b < b_{\rm max}$ with $b_{\rm min}=8$ fm and $b_{\rm max}=14$ fm for the peripheral collisions. The solid curves in Fig. 2 show the predicted isotope distributions from the ImQMD model but filtered with the impact parameter range $8 <  b < 14$ rather than the angular range. One sees that the results of ImQMD from two different filters are comparable with each other. The results of ImQMD filtered with impact parameter look slightly better at neutron-rich side comparing with the DIT predictions except the results of element $Z=35$. It seems that the peripheral collisions (i.e. larger impact parameters) have a connection with the production of more neutron-rich isotopes. In \cite{Wata}, it was also found that the most neutron-rich isotopes are connected with the low total kinectic energy loss (TKEL) part of the transfer yields. It is known that the low TKEL is correlated with the large impact parameters, which will be illustrated later.

To understand the difference between the two filters, we firstly investigate the reaction mechanism in $^{86}$Kr+$^{64}$Ni at 25 MeV/nucleon from central collisions to peripheral ones. It is expected that at incident energies around the Coulomb barrier (about $5-8$ MeV/nucleon) the reaction type in such a system is fusion or binary scattering. At Fermi energies, e.g. $E=35$ MeV/nucleon, the multi-fragmentation is expected in central and mid-central collisions. At the incident energy of 25 MeV/nucleon, the competition among fusion, multi-fragmentation and deep-inelastic scattering could appear in central and mid-central collisions. Fig. 3 shows the probabilities of fusion, binary scattering (quasi-elastic and deep inelastic scattering) and other cases (ternary breakup and multi-fragmentation of the composite system) as a function of impact parameter. From Fig. 3, we find that in central collisions the probability of fusion events is about 0.46 which is comparable with that of binary scattering events. Here, the fusion event means that the nearly spherical composite system is formed, and the channels of both evaporation residue and fusion-fission are included. Of course, the evaporation of nucleons and light fragments is simultaneously observed in the ImQMD simulations in central and mid-central collisions, due to relatively high excitation energies of the composite system. The probability of ternary breakup and multi-fragmentation is only about 0.1 in central collisions. With the increase of impact parameter, the fusion probability approaches to zero gradually whereas the probability of binary scattering events increases rapidly in semi-peripheral collisions ($b=5\sim 8$ fm). The probability of ternary breakup and multi-fragmentation goes up to 0.26 at $b=6 \sim 7$ fm. In peripheral collisions ($b > 8$ fm), the binary scattering events are dominant.

\begin{figure}
\includegraphics[angle=0,width=0.65\textwidth]{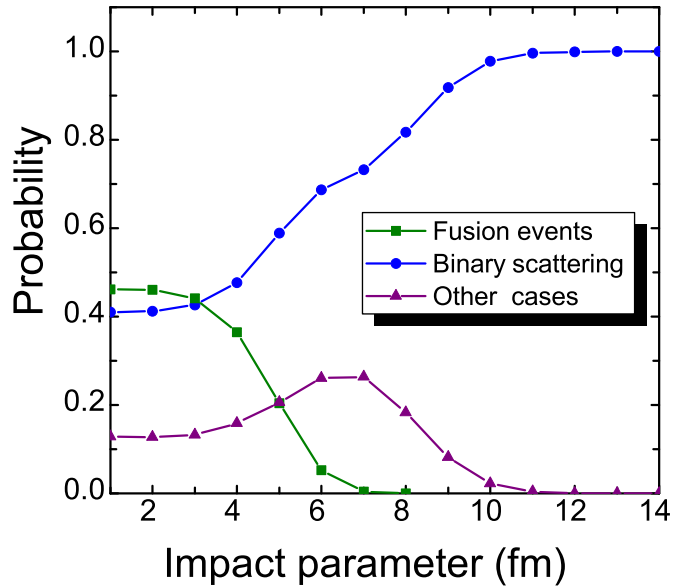}
\caption{(Color online) Probability of fusion, binary scattering and other cases as a function of impact parameter.}
\end{figure}

\begin{figure}
\includegraphics[angle=0,width=0.65\textwidth]{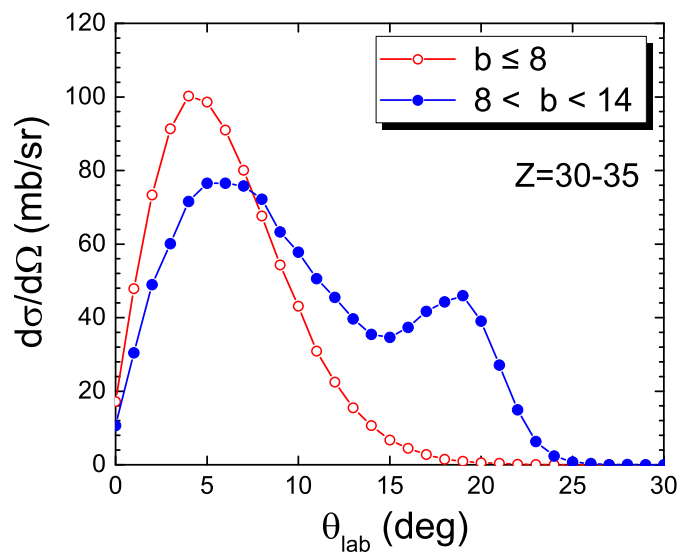}
\caption{(Color online) Differential cross sections of elements with $Z=30-35$. The open circles and solid circles denotes the results with the impact parameters $b \le 8$ fm and those with $8 <  b < 14$ fm, respectively. }
\end{figure}

\begin{figure}
\includegraphics[angle=0,width=0.65\textwidth]{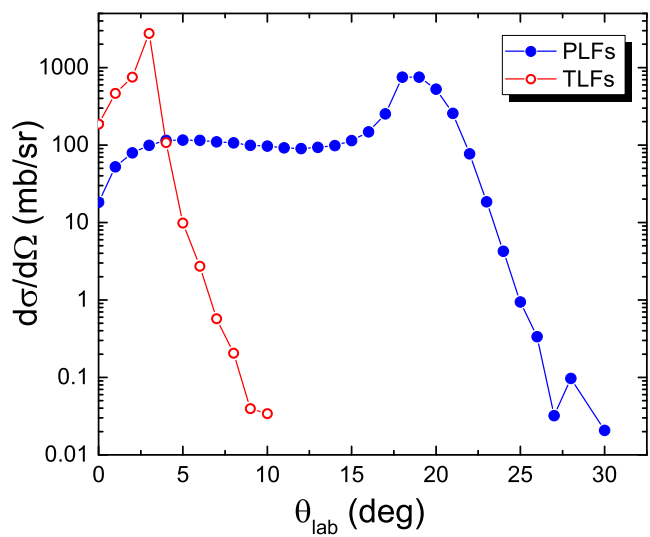}
\caption{(Color online) Differential cross sections of projectile-like fragments (PLFs) and target-like fragments (TLFs).}
\end{figure}

Fig. 4 shows the differential cross sections of elements with $Z=30-35$. The open and the solid circles denotes the results with $b \le b_{\rm min}$ and those with $b_{\rm min} <  b < b_{\rm max}$, respectively. One sees that in the angular range $10^\circ \sim 27^\circ$ the contributions from the peripheral collisions are dominant, which explains why the results from two different filters (i.e. $8< b <14$ fm and $10-27$ degree) are comparable with each other.

Simultaneously, we investigate the angular distributions of projectile-like fragments (PLFs) and those of target-like fragments (TLFs). Fig. 5 shows the differential cross sections of PLFs and TLFs. The solid and the open circles denote the results of PLFs and TLFs, respectively. One sees that in the angular range $10^\circ \sim 30^\circ$ the PLFs are dominant and the TLFs mainly locate in $\theta_{\rm lab} < 10^\circ$. It means that the measured yields using the MARS recoil separator are almost all from PLFs rather than TLFs, which is consistent with the fact that the experiment design for this reaction focus on the measurement of the cross sections and velocity distributions of projectile-like fragments \cite{Sou02}.

\begin{figure}
\includegraphics[angle=0,width=0.65 \textwidth]{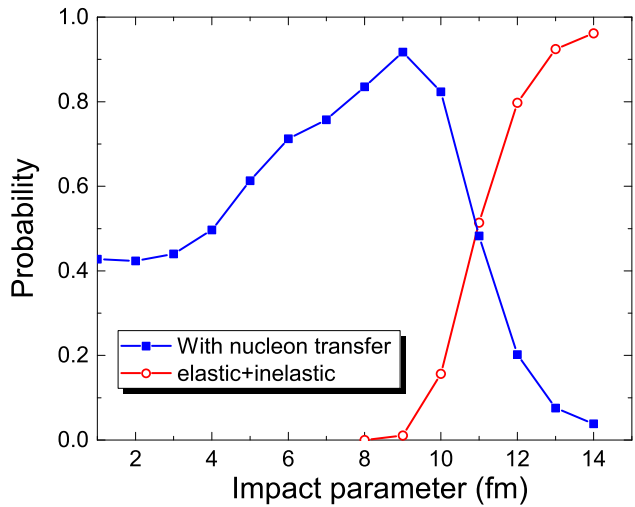}
\caption{(Color online) Probability of binary scattering events as a function of impact parameter. The squares and circles denote the results for the events with nucleon transfer and those for the elastic+inelastic events, respectively.}
\end{figure}

To explore the correlation between nucleon transfer and impact parameter in the binary scattering events, we distinguish the events with nucleon transfer between the projectile and the target nuclei from the cases elastic+inelastic scattering in semi-peripheral and peripheral collisions. Fig. 6 shows the probability of binary events as a function of impact parameter. The squares and the circles denote the results for the events with nucleon transfer and those for elastic+inelastic events, respectively. One sees that at peripheral collisions $b > 8 $ fm the probabilities of binary elastic+inelastic scattering events increase rapidly whereas those with nucleon transfer decrease sharply with impact parameter. The crossing point between the two curves is located at the touching distance $R_{12}$.

\begin{figure}
\includegraphics[angle=0,width=0.7 \textwidth]{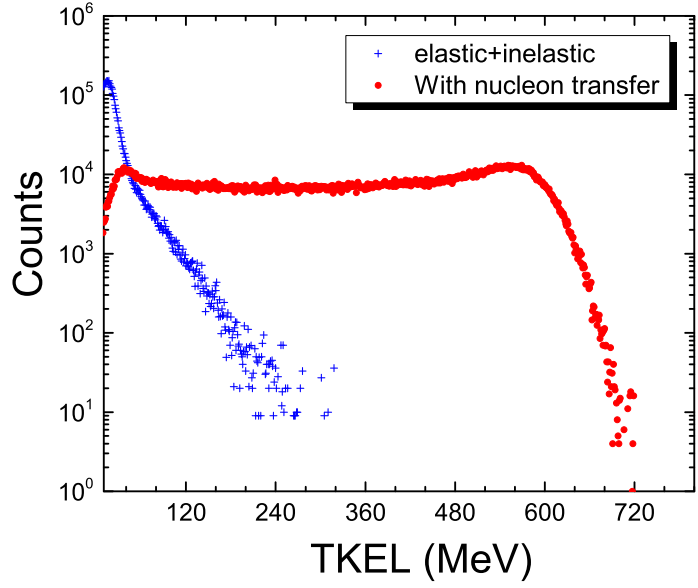}
\caption{(Color online) Distribution of TKEL of fragments in binary scattering events. The crosses and the circles denote the results for the elastic+inelastic scattering events and those with nucleon transfer, respectively.  }
\end{figure}

In addition, we investigate the influence of nucleon transfer on total kinetic energy (TKE) of outgoing fragments which is expressed as
\begin{equation}
{\rm TKE}=E_{\rm tot}-\sum_{k} E_{\rm frag}^{(k)}.
\end{equation}
Where, $E_{\rm tot}=E_{c.m.}+  E_{g.s.}^{(1)} + E_{g.s.}^{(2)}$ denotes the total energy of the reaction system at initial time.  $E_{c.m.}$ is the incident center-of-mass energy.  $E_{g.s.}^{(1)}$ and $E_{g.s.}^{(2)}$ are the ground state energy of the projectile and that of the target nucleus, respectively. $E_{\rm frag}$ denotes the energy of an outgoing fragment in its center-of-mass frame, which is expressed as
\begin{equation}
 E_{\rm frag} = \sum_i \frac{(\mathbf{p}_i-\mathbf{p}_c)^2}{2m} + U.
\end{equation}
Where, $\mathbf{p}_c$ and $U$ denote the collective momentum and the interaction potential energy of a fragment, respectively. With TKE obtained, the total kinetic energy loss (TKEL) of fragments
\begin{equation}
{\rm TKEL}=E_{c.m.}-{\rm TKE}
\end{equation}
can also be studied. Fig. 7 shows the distribution of TKEL of fragments in binary scattering events. The crosses and the circles denote the results for the elastic+inelastic events  and those with nucleon transfer, respectively. One can see evidently from the figure that the cases with nucleon transfer are distributed in a much broader region comparing with the elastic+inelastic events. We observe that most of the events with nucleon transfer are localized at TKEL values above 40 MeV. The similar platform in the distribution of TKEL was also previously observed in the strongly damped collisions of $^{136}$Xe+$^{208}$Pb, see Fig. 6 in \cite{Koz12}. It indicates that TKE of fragments is a sensitive quantity to distinguish the cases with nucleon transfer from the others.

\begin{figure}
\includegraphics[angle=0,width=0.65 \textwidth]{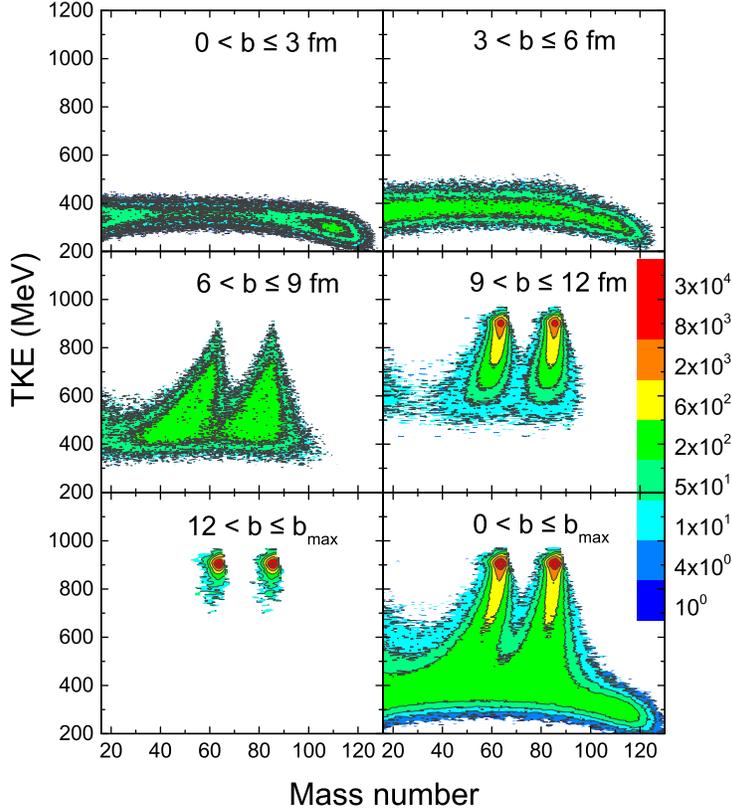}
\caption{(Color online) Mass-TKE distribution in the ImQMD calculations.}
\end{figure}

In Fig. 8, we show the mass-TKE distribution in the ImQMD calculations. One can see that the mass-TKE distribution in central collisions is quite different from those in peripheral collisions. In central collisions, the neck of the di-nuclear system can be well formed and quickly broadened at such an incident energy and many nucleons are transferred between projectile and target. As a consequence, the relative motion kinetic energy of two colliding nuclei are significantly dissipated to the excitation energy of the composite system. In addition, the masses of fragments are distributed in a much broader region, which is due to that in central collisions the yields come from quite different reaction types, such as fusion, binary breakup, ternary breakup and as well as multi-fragmentation. With the increase of impact parameter, the probability of binary scattering events increases as shown in Fig. 3. Simultaneously, the probability of elastic+inelastic scattering events rapidly increases with impact parameter in the peripheral collisions as shown in Fig. 6. The decrease of nucleon transfer results in a relatively narrow mass-TKE distribution in peripheral collision.

\begin{figure}
\includegraphics[angle=0,width=0.8 \textwidth]{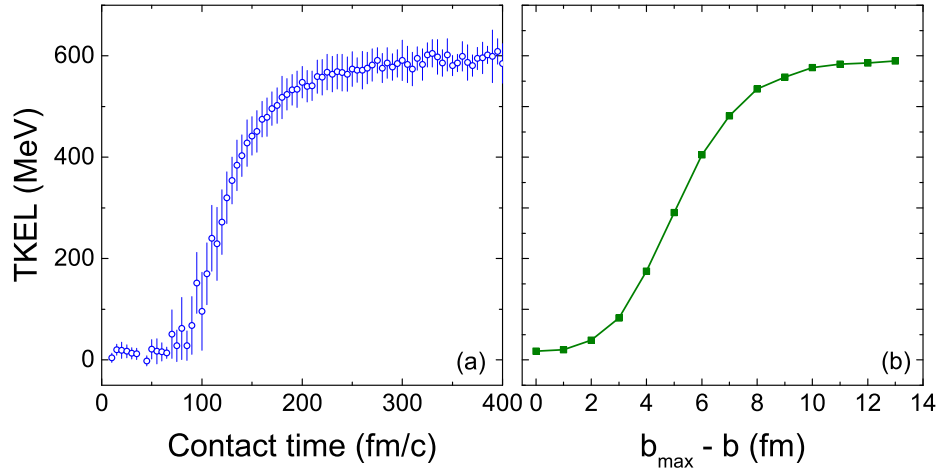}
\caption{(Color online) Predicted TKEL of system as a function of contact time (a) and as a function of impact parameter (b), here $b_{\rm max}=14$ fm.}
\end{figure}

To further understand the influence of nucleon transfer on the TKE of fragments, we investigate the contact time of the colliding system at different impact parameter. In this work, the contact time of a composite system is defined as the interval from the time at touching configuration to the time outgoing fragments being well separated. Fig. 9(a) shows the calculated TKEL as a function of reaction time. In Fig. 9(b), the correlation between the TKEL and the impact parameter is also presented for comparison. One can see that in very peripheral collisions the contact time is short due to rare nucleon transfer, and the TKEL is therefore small (namely, the low TKEL is connected with the
large impact parameters, as mentioned previously). With the decrease of impact parameter, the corresponding TKEL increases sharply, and the contact time increases simultaneously due to the increasing of nucleon transfer between projectile and target. In central collisions, the contact time reaches a few hundreds fm/c while the TKEL gradually approaches to a value of about 600 MeV.

\begin{center}
\textbf{V. SUMMARY}
\end{center}

In this work, we apply a microscopic dynamics model for description of the multi-nucleon transfer in $^{86}$Kr+$^{64}$Ni at 25 MeV/nucleon from central collisions to peripheral ones. The measured isotopic distributions for elements $Z=30-35$ can be reasonably well reproduced with the improved quantum molecular dynamics (ImQMD) model together with the statistical decay model (GEMINI) for describing the secondary decay of fragments, under two different filters: the filter with angular range $10^\circ \sim 27^\circ$ and the one from impact parameter range $8 < b < 14$ for the peripheral collisions.  We find that fusion, deep-inelastic scattering and multi-fragmentation face a strong competition in central collisions at such an incident energy. In semi-peripheral collisions, deep-inelastic scattering is dominant, and the probability of binary elastic+inelastic scattering events increases rapidly with impact parameter in peripheral collisions and approaches to one when $b>14$ fm. The predictions for the isotopic distribution indicate that the peripheral collisions (i.e. larger impact parameters) have a connection with the production of more neutron-rich isotopes. In very peripheral collisions the contact time between projectile and target is smaller than 100 fm/c in general due to rare nucleon transfer, and the corresponding TKEL is smaller than 20 MeV. Recently, the transfer form
factors up to a distance of 15.5 fm for one- and two-neutron transfers in $^{40}$Ca+$^{96}$Zr and $^{60}$Ni+$^{116}$Sn have been
measured at energies far below the Coulomb barrier \cite{Corr11,Mont14}. It would be interesting to use the ImQMD model to study, in the future, the transfer process at very large impact parameters to learn about correlations.

The distribution of total kinetic energy loss (TKEL) for binary events with nucleon transfer between projectile and target is much more broader than those of the elastic+inelastic events, which could be a sensitive quantity to distinguish the two cases from each other. The mass-TKE distribution in central collisions is quite different from those in peripheral collisions. In central collisions, the relative motion kinetic energy of two colliding nuclei are significantly dissipated to excitation energy of the composite system, and the mass number of fragments are distributed in a much broader region due to that the fragments are produced from quite different reaction types, such as fusion, binary breakup, ternary breakup and as well as multi-fragmentation. In addition, the number of nucleon transfer between projectile and target strongly influences the contact time and the TKEL. The microscopic dynamics simulations of multi-nucleon transfer (MNT) between heavy-ions not only provide a powerful theoretical approach to understand the reaction mechanism of measured colliding system, but also are useful for estimating the production cross sections and the optimal projectile-target-energy combination in the synthesis of new neutron-rich isotopes through MNT reactions.

\begin{center}
\textbf{ACKNOWLEDGEMENTS}
\end{center}
This work was supported by National Natural Science Foundation of
China (Nos. 11275052, 11365005, 11422548, 11547307, 11647309) and Guangxi Natural Science Foundation (No. 2015GXNSFDA139004). We thank an anonymous referee for valuable comments and suggestions. The executable programs of the ImQMD model are available at [http://www.imqmd.com/code/].


\begin{thebibliography}{99}


\bibitem{Li14} B. A. Li, $\rm \grave{A}$. Ramos, G. Verde, I. Vida\~{n}a, Eur. Phys. J. A\textbf{50}, 9 (2014); and references therein.

\bibitem{Stein} A. W. Steiner and S. Gandolfi, Phys. Rev. Lett. \textbf{108}, 081102
(2012).

\bibitem{Khan12} E. Khan, J. Margueron, and I. Vida\~na, Phys. Rev. Lett. \textbf{109}, 092501 (2012).

\bibitem{Tsang09} M. B. Tsang, Y. X. Zhang, P. Danielewicz, M. Famiano, Z. Li,
W. G. Lynch, and A. W. Steiner, Phys. Rev. Lett.  {\bf 102}, 122701 (2009).

\bibitem{Cent09} M. Centelles, X. Roca-Maza, X. Vinas, and M. Warda,  Phys. Rev. Lett.  {\bf 102}, 122502 (2009).

\bibitem{Wang13} N. Wang, L. Ou, and M. Liu, Phys. Rev. C \textbf{87}, 034327 (2013).


\bibitem{OSm}  J. R. Leigh, M. Dasgupta, D. J. Hinde, et al., Phys. Rev. C \textbf{52}, 3151 (1995).

\bibitem{Timm}H. Timmers, D. Ackermann, S. Beghini, \emph{et al.}, and N. Rowley, Nucl. Phys. A \textbf{633}, 421 (1998).

\bibitem{Zhang10} H. Q. Zhang, C. J. Lin, F. Yang, H. M. Jia, et al., Phys. Rev. C \textbf{82}, 054609 (2010).

\bibitem{Setf06} A. M. Stefanini, F. Scarlassara, S. Beghini et al., Phys. Rev. C \textbf{73} (2006) 034606.


\bibitem{Temis15} J. Mendoza-Temis, M. Wu, K. Langanke, G. Mart¨ªnez-Pinedo, et al., Phys. Rev. C \textbf{92}, 055805 (2015).

\bibitem{Mum16} M. R. Mumpower, R. Surman, G. C. McLaughlin, A. Aprahamian, Prog. Part. Nucl. Phys. \textbf{86}, 86 (2016).

\bibitem{LZ12} Z. Li, Z. M. Niu, B. H. Sun, et al., Acta Phys. Sin \textbf{61}, 072601 (2012).



\bibitem{OPb2} M. Dasgupta, D. J. Hinde, A. Diaz-Torres, et al., Phys. Rev. Lett. \textbf{99}, 192701 (2007).

\bibitem{Hof00} S. Hofmann and G. M\"unzenberg, Rev. Mod. Phys. \textbf{72}, 733
(2000).

\bibitem{Ogan15} Y. T. Oganessian, V. K. Utyonkov, Nucl. Phys. A \textbf{944}, 62 (2015).


\bibitem{Sob} A. Sobiczewski,  K. Pomorski, Prog. Part. Nucl. Phys. \textbf{58},
292 (2007).

\bibitem{Gup05} Raj K. Gupta, M. Manhas, G. M\"unzenberg, and W. Greiner, Phys. Rev. C \textbf{72}, 014607 (2005).


\bibitem{Wang09} N. Wang, M. Liu, Y. X. Yang, Sci. China Ser. G - Phys. Mech. Astron. \textbf{52}, 1554 (2009).

\bibitem{Wangbin} B. Wang, K. Wen, W. J. Zhao, E. G. Zhao, and S. G. Zhou, At. Data Nucl. Data Tables (in press); arXiv:1504.00756.




\bibitem{Ober14} V.E. Oberacker, A. S. Umar, C. Simenel, Phys. Rev. C \textbf{90}, 054605 (2014).
\bibitem{Shen87} W. Q. Shen, J. Albinski, A. Gobbi, et al., Phys. Rev. C \textbf{36}, 115 (1987).
\bibitem{Zag11} V. I. Zagrebaev and W. Greiner, Phys. Rev. C \textbf{83}, 044618 (2011).
\bibitem{Sou02} G.A. Souliotis, M. Veselsky, G. Chubarian, et al., Phys. Lett. B \textbf{543} (2002) 163.
\bibitem{Sou03} G. A. Souliotis, M. Veselsky, G. Chubarian, et al., Phys. Rev. Lett. \textbf{91}, 022701 (2003).
\bibitem{Heinz14} S. Heinz  and O. Beliuskina, J. Phys: Conf. Seri. \textbf{515}, 012007 (2014).


\bibitem{Wata} Y. X. Watanabe, Y. H. Kim, S. C. Jeong, et al., Phys. Rev. Lett. \textbf{115}, 172503 (2015).

\bibitem{Barr} J. S. Barrett, W. Loveland, R. Yanez, et al., Phys. Rev. C 91, 064615 (2015).

\bibitem{Li16} C. Li, et al., Phys. Rev. C \textbf{93}, 014618 (2016).

\bibitem{SmGd} N. Wang and L. Guo, Phys. Lett. B \textbf{760} (2016) 236.

\bibitem{NiU99} L. Corradi, A. M. Stefanini, C. J. Lin, S. Beghini, G. Montagnoli, F. Scarlassara, G. Pollarolo, and A. Winther, Phys. Rev. C \textbf{59}, 261 (1999).

\bibitem{Sek16} K. Sekizawa and K. Yabana, Phys. Rev. C \textbf{93}, 054616 (2016).


\bibitem{CaPb05} S. Szilner, L. Corradi, G. Pollarolo, S. Beghini, B. R. Behera, et al., Phys. Rev. C \textbf{71}, 044610 (2005).

\bibitem{CaU12} K. Nishio, S. Mitsuoka, I. Nishinaka, H. Makii, Y. Wakabayashi, H. Ikezoe, K. Hirose, T. Ohtsuki, Y. Aritomo, and S. Hofmann, Phys. Rev. C \textbf{86}, 034608 (2012).



\bibitem{UU09} C. Golabek, A. C. C. Villari, S. Heinz et al., Int. J. Mod. Phys. E \textbf{17}, 2235 (2008).
\bibitem{UU13} J. V. Kratz, M. Sch\"adel, and H. W. G\"aggeler, Phys. Rev. C \textbf{88}, 054615 (2013).
\bibitem{UUTDHF} C. Golabek and C. Simenel, Phys. Rev. Lett. \textbf{103}, 042701 (2009).
\bibitem{UUWang} N. Wang, Z. Li, and X. Wu et al., Mod. Phys. Lett. A \textbf{20}, 2619
(2005).
\bibitem{UUZhao} K. Zhao, Z. Li, X. Wu, and Y. Zhang, Phys. Rev. C \textbf{88}, 044605 (2013).

\bibitem{Zhao16} K. Zhao, Z. Li, et al., Phys. Rev. C 94, 024601 (2016).


\bibitem{Win94} A. Winther, Nucl. Phys. A \textbf{572}, 191 (1994).

\bibitem{Love15} R. Yanez and W. Loveland, Phys. Rev. C \textbf{91}, 044608 (2015).

\bibitem{Corr09} L. Corradi, G. Pollarolo, and S. Szilner, J. Phys. G: Nucl. Part.
Phys. \textbf{36}, 113101 (2009).

\bibitem{Zag08} V. Zagrebaev  and W. Greiner, Phys. Rev. Lett. \textbf{101}, 122701 (2008);  J. Phys. G: Nucl. Part. Phys. \textbf{34}, 1 (2007).

\bibitem{Ada10} G. G. Adamian, N. V. Antonenko, V. V. Sargsyan, and W. Scheid, Phys. Rev. C \textbf{81}, 024604 (2010).

\bibitem{Ada10a} G. G. Adamian, N. V. Antonenko, and D. Lacroix, Phys. Rev. C \textbf{82}, 064611 (2010).
\bibitem{Wangnan} Nan Wang, E. G. Zhao, W. Scheid, and S. G. Zhou, Phys. Rev. C \textbf{85}, 041601(R) (2012).

\bibitem{Naka05} T. Nakatsukasa and K. Yabana, Phys. Rev. C \textbf{71}, 024301 (2005).

\bibitem{Maru06} J. A. Maruhn, P.-G. Reinhard, P. D. Stevenson, and M. R. Strayer, Phys. Rev. C \textbf{74}, 027601 (2006).

\bibitem{Guo07} Lu Guo, J. A. Maruhn, and P.-G. Reinhard, Phys. Rev. C \textbf{76}, 014601 (2007).

\bibitem{Sim12} C. Simenel, Eur. Phys. J. A \textbf{48}, 152 (2012).


\bibitem{Sim14} C. Simenel, J. Phys. G: Nucl. Part. Phys. \textbf{41}, 094007 (2014); arXiv:1403.3246v1.




\bibitem{ImQMD2002} N. Wang, Z. X. Li, and X. Z. Wu, Phys. Rev. C \textbf{65}, 064608 (2002).

\bibitem{ImQMD2004} N. Wang, Z. X. Li, X. Z. Wu, J. L. Tian, et al., Phys.
Rev. C \textbf{69}, 034608 (2004).

\bibitem{ImQMD2010} Y. Y. Jiang, N. Wang, Z. X. Li and W. Scheid, Phys. Rev. C \textbf{81}, 044602 (2010).

\bibitem{ImQMD2014} N. Wang, L. Ou, Y. X. Zhang and Z. X. Li, Phys. Rev. C \textbf{89}, 064601 (2014).


\bibitem{constrain} M. Papa, T. Maruyama, and A. Bonasera, Phys. Rev. C \textbf{64}, 024612 (2001).


\bibitem{Wang15} N. Wang, K. Zhao, Z. X. Li, Sci. China-Phys. Mech. Astron. \textbf{58}, 112001 (2015).
\bibitem{Wang16} N. Wang, T. Wu, J. Zeng, et al., J. Phys. G: Nucl. Part. Phys. \textbf{43}, 065101 (2016).



\bibitem{Char88} R. Charity, et al., Nucl. Phys. A \textbf{483}, 371 (1988).

\bibitem{Wang14} N. Wang, M. Liu, X. Z. Wu, and J. Meng, Phys. Lett. B \textbf{734}, 215 (2014); http://www.imqmd.com/mass/WS4.txt


\bibitem{Tas91} L. Tassan-Got, C. Stefan, Nucl. Phys. A \textbf{524}, 121(1991).

\bibitem{Radii} N. Wang and T. Li, Phys. Rev. C \textbf{88}, 011301(R) (2013).

\bibitem{Koz12} E. M. Kozulin,  E. Vardaci,  G. N. Knyazheva, et al., Phys. Rev. C \textbf{86}, 044611 (2012).

\bibitem{Corr11} L. Corradi, S. Szilner, G. Pollarolo, et al., Phys. Rev. C \textbf{84}, 034603 (2011).

\bibitem{Mont14} D. Montanari,  L. Corradi,  S. Szilner, et al., Phys. Rev. Lett. \textbf{113}, 052501 (2014)

\end{thebibliography}
\end{document}